\documentclass[12pt]{iopart} \usepackage{epsfig}

\newcommand{\fvec}{{\bf F}}

\newcommand{\bnabvin}{{\bfnabla_{{\bf v}_i}}}

\newcommand{\be}{\begin{equation}}
\newcommand{\ee}{\end{equation}}
\newcommand{\bea}{\begin{eqnarray}}
\newcommand{\eea}{\end{eqnarray}}
\newcommand{\barr}{\begin{array}}
\newcommand{\earr}{\end{array}}

\newcommand{\bfxi}{\mbox{\boldmath $\xi$}}

\newcommand{\xx}{{\bf x}}
\newcommand{\rr}{{\bf r}}
\newcommand{\NN}{{\bf \nabla}}
\newcommand{\FF}{{\bf F}}

\newcommand{\vv}{{\bf v}}
\newcommand{\cc}{{\bf c}}
\newcommand{\uu}{{\bf u}}

\renewcommand{\ss}{{\bf\sigma}}

\newcommand{\shat} {\hat{\ss}}

\newcommand{\bfnabla}{\mbox{\boldmath $\nabla$}}
\newcommand{\bnabri}{{\bfnabla_{{\bf r}_i}}}

\usepackage{iopams}  
\usepackage{setstack}
\begin{document}

\title{Dynamic density functional theory
versus Kinetic theory of simple fluids }

\author{Umberto Marini Bettolo Marconi\footnote[3]
{(umberto.marinibettolo@unicam.it)}
}

\address{Dipartimento di Fisica, Universit\`a di Camerino and Istituto
Nazionale di Fisica della Materia, Via Madonna delle Carceri, 62032 ,
Camerino, Italy, INFN, sez. di Perugia, Italy }
 
\author{Simone Melchionna}

\address{Institute of Materials, Ecole Polytechnique F\'ed\'erale de 
Lausanne (EPFL), 
1015 Lausanne, Switzerland}

\begin{abstract}
By combining methods of kinetic  and density functional theory,
we present a description of molecular fluids
which accounts for their  microscopic structure and thermodynamic
properties as well as for the hydrodynamic behavior.
We focus on the evolution of the one particle phase space 
distribution, rather than on the evolution of the average particle density, 
which features in dynamic density functional theory. 
The resulting equation  can be studied 
in two different physical limits: 
diffusive dynamics, typical of
colloidal fluids without hydrodynamic interaction,  where particles 
are subject to overdamped motion resulting from the coupling with a 
solvent at rest,  and inertial dynamics, typical of
molecular fluids.
Finally, we propose an algorithm to solve
numerically and efficiently the resulting kinetic equation by
employing a discretization procedure analogous to the one used in the
Lattice Boltzmann method.
 
\end{abstract}

\section{Introduction}
\label{intro} 

Understanding  transport phenomena and their interplay with  structural
properties  of liquids  in strongly inhomogeneous systems is of capital importance
in many fields such as microfluidics, science of materials,  physiology and 
has several technological applications,
ranging from pharmaceuticals to ternary oil recovery \cite{Squires}.
Much progress has been achieved in the last thirty  years concerning the equilibrium 
properties of liquids in the presence of confining geometries 
or  structured substrates.
One of the major tools to explore
 inhomogeneous 
classical fluids
 is the Density Functional Theory
(DFT), yielding thermodynamic  (free energy, phase coexistence, osmotic pressure)
and  structural properties (density and composition profiles) within the same framework ~\cite{Evans1,Evans2}.
Such a method presents many advantages: it is highly
versatile, variational, lends itself to approximations guided by physical insight and
 is universally applicable  to the 
study the properties of all real fluids under a
wide variety of thermodynamic and geometric conditions.
In view of the great success of the DFT, a Dynamic DFT  (DDFT) has been proposed to describe the dynamics
of the density profile $n(\rr,t)$ possibly due to a time dependent external potential or to 
a sudden change of the control parameters \cite{Tarazona,ArcherEvans}. 
%There is a consensus, that
%under appropriate conditions such as those realized in a colloid-solvent
%solution the DDFT   
%may provide a valid description of the approach towards  equilibrium. 

The typical DDFT  equation is a continuity equation for the 
number density:
\begin{eqnarray}
\frac{\partial {n} (\rr,t)}{\partial t}+\nabla \cdot {\bf J}(\rr,t)=0
\label{uno}
\end{eqnarray}
supplemented by a prescription relating the particle flux, ${\bf J}$,
to the density itself obtained by using the specific properties of the
dynamics of the system. In the case of a colloid-solvent system without hydrodynamic interaction,
one ideally  traces out the solvent  microscopic
degrees of freedom and mimics their influence on the solute particles 
via  Brownian dynamics induced by a heat bath stochastic term.
Under this assumption  one can derive a simple
relation between ${\bf J}$ and $n$. On the other hand,
in the case of molecular fluids, the absence 
of the fast degrees of freedom of the solvent seems to preclude
such a possibility. It is necessary to recognize that the minimal set
of collective variables which are relevant in the dynamical 
description of a fluid corresponds to the locally conserved fields,  namely  
number of particles, momentum  and energy density.
%In the absence of a solvent these vaiables evolve slowly. 
One can represent their evolution at a mesoscopic level 
by means of a set of hydrodynamic equations,
which   require the specification of 
the currents associated with momentum and the energy densities
in terms of the hydrodynamic variables via 
the so called phenomenological constitutive relations.
However, if one requires a microscopic level of description similar to the one
achieved by DDFT it is necessary to choose a kinetic approach and
consider
the time evolution of 
the one-particle phase space distribution, $f(\rr,\vv,t)$.
Doing so it is also possible to
bridge hydrodynamics and microscopic
structural theories of fluids. 
The method  avoids the difficulty of computing separately
the velocity moments of the distribution function because it 
considers directly the evolution of $f$. The derived equations can then be solved numerically
by means of the Lattice Boltzmann method, a well-known method widely used in computational fluid dynamics \cite{LBgeneral}.
The resulting theory is able to
predict structure and
thermodynamics with an accuracy comparable to that of DDFT, 
but has the advantage of 
describing  transport properties more realistically \cite{Melchionna2008,Melchionna2009}.
In addition we shall show that from the present kinetic approach 
it is possible to derive both the DDFT for colloids
 and the transport equation describing
molecular fluids. The two descriptions depart when one specifies the 
interaction
of the fluid with the heat bath.
In the first case, the heat bath is fixed, breaks the Galilei invariance
of the system and damps the momentum and energy transport. In the second case,
the internal degrees of freedom act as an heat bath co-moving
with the fluid and restoring the local equilibrium.

%%%%%%%%%%%%%%%%%%%%%%%%%%%%%%%%%

The present paper is organized as follows:
in sec. \ref{colloid} we shall briefly review 
the DDFT for colloid-solvent solution by considering its derivation
from an approach based on the evolution of the 
one particle phase space distribution. In sec. \ref{molecular}
we switch to the case of a molecular liquid and 
obtain a simplified transport equation which accommodates within the same
framework thermodynamic and structural properties,
inertial effects of the fluid (sound waves, shear modes) and
transport coefficients.
In sec. \ref{LBE} we give a brief account of the numerical method
used to solve the transport equation.
% and some numerical examples.
Finally in sec. \ref{conclusions} we present
the conclusions.

\section{ Microscopic description  }
\label{colloid}
Let us consider first   a system of great practical interest,  
$N$  particles, of mass $m$ and positions $\rr_i$
and velocities $\vv_i$, immersed in a solvent.
Since the solvent particles are in general much lighter and smaller
than the host particles, 
the degrees of freedom of the solvent  can be eliminated by a
applying a suitable coarse graining procedure.
The resulting total force acting on each colloidal particle is the sum of 
the interactions with other colloidal particles, of the external fields and of 
a stochastic force represented  as a Langevin process, which combines a 
frictional force proportional to the velocity with respect to the solvent
with a stochastic white noise forcing $-\gamma^{solv}\vv_i+\bfxi_i$.
%forcing $\bfxi_i$ ($<\bfxi>=0$) stemming from collisions
%with the solvent particles.
Since both terms  are originated from the solvent  their amplitudes are related
by the following fluctuation-dissipation relation
$
\langle \xi^\alpha_{i}(t)
\xi^\beta_{j}(s) \rangle  = 2 \gamma^{solv} m k_B T^{solv}\delta_{ij}
\delta^{\alpha\beta} \delta(t-s), 
$ 
where the superscript "{\em solv}" indicates the origin of the friction and of the thermostatting medium. 
%Such a term represents an heat bath and provides the
%key ingredient to derive  
%the DDFT equation from a microscopic  model.
The
particles  interact through a pair potential $U$ and 
are subject to a force field $\fvec_{ext}$:

\begin{eqnarray}
%\label{kramers-a}
&\frac{d\rr_i}{dt} &= \vv_i  \nonumber\\
& m \frac{d \vv_i}{dt} & = \left[ 
\fvec_{ext}(\rr_i) 
- \sum_{j(\neq i)} \bnabri U(|\rr_i-\rr_j|) \right]
- m \gamma \vv_i   + \bfxi_i(t)
\label{kramers-b}
%\nonumber
\end{eqnarray}

It is straightforward to derive 
the associated linear Kramers-Fokker-Planck 
 \cite{Risken}
evolution equation for the $6N$ dimensional phase-space probability 
density distribution $f_N(\{\rr,\vv\},t)$, where $\{\rr,\vv\}$ indicates a $6N$ dimensional phase space point:
\bea
&&\Bigl(\frac{\partial}{\partial t}+\sum_i \Bigl[\vv_{i}\cdot\bnabri+[\frac{\
\fvec_{ext}(\rr_i)}{m}-\frac{1}{m} \sum_{j (\neq i)}
\bnabri U(|\xx_i-\xx_j|)]\cdot\bnabvin \Bigr]
\Bigl) f_N(\{\rr,\vv\},t) 
\nonumber \\
&=& 
\gamma^{solv}\sum_i \Bigr[\bnabvin\cdot \vv_i +\frac{k_B T^{solv}}{m}\bnabvin^2 \Bigr] f_N(\{\rr,\vv\},t) 
\label{many4}
\eea
%If  $\gamma^{solv}=0$, 
%the dynamics is conservative and equation  (\ref{many4}) 
Notice that we switched from a description in terms of trajectories of
the particles to a probabilistic representation in terms of
$f_N$. This can be justified in two possible ways: the probabilistic
description is a) the result of an ensemble averaging of the
trajectories over a noise ensemble, or b) follows from an averaging
over initial conditions.  Case a) applies to the damped stochastic
dynamics ($\gamma^{solv}>0$) typical of a colloidal suspension, while
case b) applies to the Hamiltonian dynamics characteristic of an
atomic liquid, $\gamma^{solv}=0$, when eq. (\ref{many4}) reduces to
the Liouville equation.

The information contained in $f_N$ is fully microscopic since it
describes the microstate of the system. 
However, one recognizes that after an initial many-body regime,
whose duration is  of the order of the duration of a collision event,
such a representation becomes redundant
and is possible to contract the description from $6N$ dimensions to
only $6$ dimensions, that is, from the phase space distribution of $N$
particles to the one particle phase space distribution,
$f(\rr,\vv,t)$.  In this stage, termed kinetic regime, the one-body
distribution relaxes towards a local Maxwellian and its
evolution can be represented by the following equation:

\bea
\bigl( \frac{\partial}{\partial t}  +\vv\cdot\NN 
+\frac{\fvec_{ext}(\rr)}{m}\cdot\frac{\partial}{\partial \vv}\bigr) f(\rr,\vv,t)
= {\cal Q}(\rr,\vv,t)+{\cal B}^{solv}(\rr,\vv,t).
\label{equilproc}
\eea 
This equation contains the streaming terms in the l.h.s.,
whereas in the r.h.s. ${\cal Q}$ is the interaction term 
\be
{\cal Q}(\rr,\vv,t)=
\frac{1}{m}{\NN_{\bf v}} \cdot \int d\rr'\int d\vv' 
f_2(\rr,\vv,\rr',\vv',t){\bf \NN_{\bf r}}U(|\rr-\rr'|) 
\label{bbgky}
\ee
 and ${\cal B}^{solv}$ represents the heat bath term:
\be
{\cal B}^{solv}(\rr,\vv,t)=\gamma^{solv}[\frac{k_B T^{solv}}{m}\frac{\partial^2} 
{\partial \vv^2} 
+\frac{\partial} {\partial \vv}\cdot \vv ]f(\rr,\vv,t)
\label{lfp}
\ee
To proceed further 
one makes the approximation that the two particle distribution function,
$f_2$ ,can be expressed as a product of one particle distribution functions
times the positional  pair correlation function $g$:
\be
f_2(\rr,\vv,\rr',\vv',t)\approx 
f(\rr,\vv,t) f(\rr',\vv',t) g(\rr,\rr')
\nonumber 
\label{factor}
\ee
so that eq. (\ref{equilproc}) becomes a closed non-linear equation for the one
particle distribution.
This is a crucial assumption since it decouples the evolution
of $f(\rr,\vv,t)$ from the evolution of the higher order multiparticle distribution 
functions. Many-particle correlations are however retained  through the structural information contained in
the positional  pair correlation function function $g(\rr,\rr')$. As an approximation, we shall assume that
$g(\rr,\rr')$ is a non local function of the profile $n(\rr,t)$, depends on time only
through the density profile and has the same form as in a nonuniform 
equilibrium state whose density is $n(\rr,t)$.

For $\gamma^{solv}>0$,
one can  simplify further equation (\ref{equilproc}) 
and derive an equation involving only
the ensemble averaged particle density, by using the fact 
momentum and energy of the colloidal particles are not conserved,
so that the currents become rapidly
``slaved'' to the density, i.e. the 
evolution is completely determined in terms of $n(\rr,t)$ and its derivatives.
That occurs because
the velocities of the particles rapidly relax towards the
equilibrium distribution, in a  time of order
of the inverse friction time $\tau^{solv}=1/\gamma^{solv}$. This is the case when 
the non-dimensional ratio $v_T\tau^{solv}/\sigma<< 1$
where $\sigma$ is the typical
size of the particles and $v_T$ the thermal velocity,
since the particles undergo a small displacement in a time $\sim\tau^{solv}$.
The only relevant evolution on  timescales larger than $\tau^{solv}$ regards
the spatial distribution of the particles. 
 This is the
reason why the DDFT gives a sufficiently accurate
description of colloidal systems.

To be concrete, let us define the velocity moments of $f(\rr,\vv,t)$
by multiplying eq. (\ref{equilproc}) by $1,m\vv, m(\vv-\uu)^2/2$, respectively
and integrating w.r.t. $\vv$, obtaining the continuity equation:

\be
\partial_{t}n(\rr,t) +\nabla\cdot ( n(\rr,t) \uu(\rr,t))=0
\label{continuity}
\ee 
the momentum balance equation 
\bea
\fl
m n(\rr,t)[\partial_{t}u_j(\rr,t)+u_i(\rr,t) \partial_i u_j(\rr,t)] 
+\partial_i P_{ij}^{(K)}(\rr,t) -F_j(\rr)n(\rr,t)- C_j^{(1)}(\rr,t)=
b_j^{(1)}(\rr,t)\nonumber\\
\label{momentum}
\eea
and the kinetic balance equation

\bea
\fl
\frac{3}{2}k_B n(\rr,t) [\partial_{t} +u_i(\rr,t)\partial_i] T(\rr,t)+P^{(K)}_{ij}(\rr,t) \partial_i u_j(\rr,t)
+\partial_i q^{(K)}_i(\rr,t)-C^{(2)} (\rr,t) =b^{(2)}(\rr,t)
\nonumber\\
\label{energy}
\eea
where  the local fluid density is
\be
n(\rr,t)= \int d\vv f(\rr,\vv,t)
\label{density}
\ee
the local fluid velocity, $\uu$, is
\be
n(\rr,t)\uu(\rr,t)=\int d\vv \vv f(\rr,\vv,t)
\label{velocity}
\ee
%Since each of these fields is locally conserved in the absence
%of external forces, when
%p%erturbed, it slowly decays to thermodynamic equilibrium.
and the local fluid kinetic energy density is:
\be
e^{kin}(\rr,t)=\frac{d}{2} k_B n(\rr,t) T(\rr,t)=
\frac{m}{2}\int d\vv (\vv-\uu)^2 f(\rr,\vv,t)
\label{temperature}
\ee
We have introduced the kinetic component of the pressure
tensor 
\be
P^{(K)}_{ij}(\rr,t)=m\int d\vv f(\rr,\vv,t)(\vv-\uu)_i(\vv-\uu)_j
\label{pk}
\ee
and of the  heat flux  vector:
\be
q^{(K)}_{i}(\rr,t)=\frac{m}{2}\int d\vv f(\rr,\vv,t)(\vv-\uu)^2(\vv-\uu)_i
\label{qk}\ee
Notice that $P^{(K)}_{ij}$ and $q^{(K)}_{i}$
cannot in general be expressed in terms of the hydrodynamic moments 
introduced so far.
We also have  two terms stemming from the interaction

\be
\fl
C_i^{(1)}(\rr,t)  =m\int d\vv   {\cal Q}(\rr,\vv,t) v_i
= -\frac{1}{m}n(\rr,t)\nabla_i\frac{\delta{\cal F}^{nonideal}[n(\rr,t) ]}
{\delta n(\rr,t)}
= -\partial_j  P_{ij}^{(C)}(\rr,t)
\nonumber\\
\label{ci}
\ee

\be
\fl
C^{(2)} (\rr,t) =\frac{m}{2}
\int d\vv  {\cal Q}(\rr,\vv,t)  (\vv-\uu)^2
\label{c2}
\ee
The second equality in (\ref{ci}) relates $C_i^{(1)}$
to a functional derivative of the non ideal part of the free energy,
whereas the third equality connects it to a
spatial derivative of the
potential part of the stress tensor.
It is important to notice that $C_i^{(1)}$ and $C^{(2)}$
 vanish in a uniform system. 
Finally we specify the terms stemming the coupling with the heat bath
\bea
\fl
b_i^{(1)}(\rr,t)  =m\int d\vv(v-u)_i  {\cal B}^{solv}(\rr,\vv,t)=
-m\gamma^{solv}n(\rr,t)u_i(\rr,t)
\eea

\bea
\fl
b^{(2)} (\rr,t) =\frac{m}{2}
\int d\vv   (\vv-\uu)^2 {\cal B}^{solv}(\rr,\vv,t)
=\gamma^{solv}[3 k_B T^{solv}-m<\vv^2>]n(\rr,t)
%\nonumber\\
\eea

The presence of $b_i^{(1)}, b^{(2)}$, 
in  eqs. (\ref{momentum}-\ref{energy})
determine a fast  equilibration process of the momentum current
$n\uu$
and of the local temperature $T$ towards their stationary
values.
This fact is used to contract the description from
the five fields appearing in eqs. (\ref{continuity}-\ref{energy})
to the single field appearing in the DDFT equation. The rigorous mathematical procedure to derive
this result employs the
multiple time scale analysis \cite{UMBtar,UMBcecc,UMBmel},
exploiting the
time scale separation between
the zero mode associated with the density fluctuations
and the remaining modes, including the momentum and energy
fluctuations which are fast relaxing due to the friction with the
solvent.
 Physically speaking, the  heat bath renders possible the contraction
from the phase space to the configurational space
because it rapidly  ``washes out''
any possible deviation of the distribution function from the 
Gaussian.
The larger the friction the faster the restoring process towards the 
Maxwellian.
We shall not go through the details of such a derivation which can be found
in refs. \cite{UMBtar,UMBmel} and merely quote the final result:
\be
 \partial_{t}n(\rr,t) =\frac{1}{m \gamma}\sum_i
\nabla_i [k_B T \nabla_i n(\rr,t) -F_i(\rr)n(\rr,t)- C_i^{(1)}(\rr,t)].
\ee

It is possible to compute inertial corrections  to such an equation
by 
employing systematically the multiple time scale analysis,
which is tantamount of an inverse friction expansion, but this procedure does not
help in the case of molecular fluids where the inverse friction parameter
$1/\gamma^{solv}$ diverges, and only internal dissipation mechanisms are at work.
A salient feature of molecular liquids  is that they  support
hydrodynamical modes, as a result of local
conservation laws of particle number, momentum and energy,  which are eventually damped 
due to internal friction and heat transport mechanisms.
Hence it is necessary to apply a  strategy 
which preserves the translational invariance of the system
and does not select a particular reference frame where the solvent is at rest \cite{Espanol,Archer2009}.

%%%%%%%%%%%%%%%%%%%%%%%%%%%%%%%%%%

\section{ Molecular fluids  }
\label{molecular}

The key difference between colloidal systems
and molecular fluids is that in the former the equilibration
is due to the heat bath externally imposed by the solvent
whilst in the latter equilibration is realized via
viscosity and heat conduction among the same fluid elements.
In molecular fluids typical relaxation times of hydrodynamic modes
diverge as $k^{-2}$ for excitations of wave vector $k\to 0$,
whereas the relaxation times of non-hydrodynamic modes are shorter and finite
in the same limit. Our treatment of 
eq.(\ref{equilproc}) must retain such a feature while accounting for
the structural and thermodynamic properties with an accuracy comparable
with DDFT \cite{Brey}.
%the heat bath  externally imposed by solvent, whereas
%in
%molecular fluids the equilibration is determined by the fluid  itself, 
%via viscosity and heat resistivity.
%However, the central question is:  can we export DDFT concepts to 
%the study atomic liquids?
%We remark that
%the  first $2+d$ moments of $f(\rr,\vv,t)$  have a privileged status
%featuring in eqs.  \ref{continuity}-\ref{energy}, since
%they are conserved fields in the absence of the solvent friction. 
%The remaining moments instead decay rapidly because are not associated with 
%conservation laws. They
% absorb energy from the soft modes and restore global equilibrium. 
For $\gamma^{solv}=0$, we recast eq.(\ref{equilproc}) 
 as:
\begin{eqnarray}
\fl
\partial_{t}f(\rr,\vv,t) +\vv\cdot\NN f(\rr,\vv,t)
+\frac{\FF^{ext}(\rr)}{m}\cdot\frac{\partial}{\partial \vv} f(\rr,\vv,t)=
%\nonumber\\ &&
{\cal K}(\rr,\vv,t)+{\cal B}^{int}(\rr,\vv,t)\nonumber\\
\label{unobrey}
\end{eqnarray}
where
\be
\fl
{\cal K}(\rr,\vv,t)\equiv
\frac{ f_{loc}(\rr,\vv,t)}{k_Bn(\rr,t) T(\rr,t)}
\Bigl((\vv-\uu)\cdot
{\bf C}^{(1)}(\rr,t)
+(\frac{ m(\vv-\uu)^2}{3 k_B T(\rr,t)} -1 )C^{(2)}(\rr,t)\Bigl)
\ee
and
\be
f_{loc}(\rr,\vv,t)= 
n(\rr,t)[\frac{m}{2\pi k_B T(\rr,t)}]^{3/2}\exp
\Bigl(-\frac{m(\vv-\uu)^2}{2 k_B T(\rr,t)} \Bigl).
\ee
The two terms in the r.h.s. of eq.(\ref{unobrey}) represent a suitable decomposition of
${\cal Q}(\rr,\vv,t)$.
The idea 
is to  treat  the contribution of ${\cal Q}$ from 
hydrodynamic modes separately from the non-hydrodynamic kinetic modes. 
The term  $ {\cal K}$ guarantees that  the local transfer of momentum and energy is not altered.
The part ${\cal B}^{int}$ is chosen so that
it does not contribute  explicitly neither to the evolution 
of the hydrodynamic modes nor to the structure
of the fluid, i.e. 
it does not appear in eqs. (\ref{momentum}-\ref{energy}).
For this reason it can be treated within a simple relaxation approximation,
as shown below.
 We have deliberately employed the same 
symbol ${\cal B}$ used for the solvent heat-bath,
to stress the fact that even in this case a thermalizing mechanism
 for the hydrodynamic modes exists and stems from the interplay
between these modes and the non-hydrodynamic modes. 
The  role  of   ${\cal B}^{int}$ 
is to  reproduce in an approximate and simple fashion the fast
relaxation process of $f$ 
towards the local equilibrium distribution $f_{loc}$.
The following  BGK relaxation approximation \cite{BGK} for the non-hydrodynamic
components of ${\cal Q}$
meets the requirements of leaving unaltered the balance equations
\be
{\cal B}^{int}(\rr,\vv,t)\equiv -\nu_0[ f(\rr,\vv,t)-f_{loc}(\rr,\vv,t)]
=-\nu_0 \delta f(\rr,\vv,t)
\label{ansatzbgk}
\ee
where $\nu_0$ is a microscopic relaxation rate
of the order of Enskog collision rate, $\omega_E=4\sqrt\pi n\sigma^2 v_T$,
where $\sigma$ is a molecular diameter and $v_T=\sqrt{k_B T/m}$.
Since by construction both $f(\rr,\vv,t)$ and $f_{loc}(\rr,\vv,t)$  correspond 
to the same values of $n,\uu,T$, one obtains 
that $b^{(1)}_i$ and $b^{(2)}$,  associated with ${\cal B}^{int}$,
vanish.
In conclusion, the balance equations for the
hydrodynamic moments  generated by eq.(\ref{unobrey}) 
with the ansatz (\ref{ansatzbgk})
 meet all the requirements imposed by the translational
invariance of the system and thus possess the correct long-wavelength 
hydrodynamic behavior.

Before particularizing the description
we wish to comment on a similarity between the DDFT 
and the present approach. It was found in DDFT that
the free energy always decreases with time:
$$
\frac{d {\cal F} [n]}{d t}=-\int d\rr n(\rr,t) )\bigl[ \bfnabla\frac{ 
\delta {\cal F}[n]}{\delta n(\rr,t)}\bigr]^2\leq 0
.$$
In the present case instead we have that 
the entropy, defined as 
$$
S[f]=-k_B \int d\rr d\vv f(\rr,\vv,t)\ln f(\rr,\vv,t)
$$
increases with time   $ \frac{d S[f]}{d t}\geq 0$, which
is a statement of the $H$-theorem.
%%%%%%%%%%%%%%%%
%%%%%%%%%%%%%%%%cccccccccccccccccccccccccc

Using the special form  of 
${\cal B}^{int}$ it is possible  to obtain perturbatively an approximation for  
 $\delta f$ and the transport coefficients at low density.
In the spirit of the Chapman-Enskog method \cite{Huang} we insert the trial distribution $f_{loc}(\rr,\vv,t)$ in the l.h.s 
of eq. (\ref{unobrey}) and neglect the term ${\cal K}$. Using eqs.Ê(\ref{continuity}) - Ê(\ref{energy})
to eliminate the time derivatives of the fields
we arrive at 
the following explicit representation of $\delta f$
\bea
 \delta f(\rr,\vv,t)&=&-\frac{1}{\nu_0}\frac{  f_{loc}(\rr,\vv,t) }{k_B T(\rr,t)}
\Bigl[ \Bigl ( \frac{m}{2}\frac{(\vv-\uu)^2}{k_B T(\rr,t)} -
\frac{5}{2} \Bigr)(v_i-u_i)\partial_i  k_B T(\rr,t) \nonumber\\ &+& m \Bigl
( (v_i-u_i)(v_k-u_k)- \frac{(\vv-\uu)^2}{3}\delta_{ik}
\Bigr)\partial_i u_k (\rr,t)\Bigr ].
%\nonumber 
\label{perturbed}
\eea
which is valid in the low density limit and to first order in the gradients of $\uu$ and $T$.
By substituting  $\delta f$ in eqs. (\ref{pk})
,(\ref{qk}) and with the help of the definitions
$$
P_{ij}^{(K)}=-\eta^{(K)}\Bigl(\nabla_i u_j+\nabla_j u_i
-\frac{2}{3}\nabla\cdot \uu\Bigl)
$$
and
$$
q_i^{(K)}=-\lambda^{(K)}\nabla_i T
$$
we compute the associated kinetic contributions to the 
viscosity
\be
\eta^{(K)}=\frac{ m v_T^2}{\nu_0}n
\label{viscosityk}
\ee
and to  the heat  conductivity
\be
\lambda^{(K)}=\frac{5}{2}\frac{v_T^2}{ \nu_0}  k_B n.
\ee

%%%%%%%%%%%%%%%%%%%%%%%%%%%%%%%
%%%%%%%%%%%%%%%%%%%%%%%%

In order to obtain
the ``potential'' contribution to the transport coefficients one needs
 an explicit representation of $C^{(1)}_i$ and $C^{(2)}$,
which can be studied 
 either in the random phase approximation  (RPA) \cite{hansen} consisting
in combining eqs. (\ref{bbgky}) and (\ref{factor})
or in the  Revised Enskog Theory (RET), which applies to hard sphere fluids
and to piecewise constant potentials.

Within the RPA one finds  $C^{(2)}(\rr,t)=0$ and

\be
C_i^{(1)}(\rr,t) ={n(\rr ,t)} F_i^{mol}(\rr,t)
\ee
where we have introduced the self consistent molecular field:
\be
 F_i^{mol}(\rr,t)=
-\int d\rr' n(\rr',t) g(\rr,\rr')\frac{\partial } 
{\partial r_i} U(|\rr-\rr'|). 
\ee

Concerning the RPA, we wish to comment that
it can treat fluids characterized by 
slowly varying potentials, like ultra-soft potentials used to 
model the steric repulsion between polymers or long range attractive interactions. It can describe qualitatively  their multiphase behavior ,
 but does not give a suitable representation of  their transport properties in the high density region, as it is evident
 from the vanishing of $C^{(2)}(\rr,t)$.
In fact, the RPA treatment of correlations while accounting for
thermodynamical properties, does not contribute to their
hydrodynamical properties.

Sharp repulsive interactions cannot be accounted for as a molecular field
because they imply strong correlations between the relative positions
and the relative velocity.

On the contrary, one can obtain a satisfactory representation of  both
structural and transport properties of hard sphere fluids.
A very accurate treatment of how the hard sphere interaction  
contributes to the
evolution of the phase space distribution is provided 
by the RET \cite{VanBeijeren}
which approximates the collisional integral as:
\bea
&&{\cal Q}_{RET}[f](\rr,\vv_1,t)
= \sigma^{d-1}\int d\vv_2\int 
d\hat{\ss}\Theta(\hat{\ss}\cdot \vv_{12}) (\hat{\ss} 
\cdot \vv_{12})\times\nonumber\\
&&\{ g_{hs}(\rr,\rr-\ss\shat) f (\rr,\vv_1',t)f (\rr-\ss\shat,\vv_2',t)  -
g_{hs}(\rr,\rr+\ss\shat)f(\rr,\vv_1,t)f(\rr+\ss\shat,\vv_2,t)\}\nonumber\\
\label{ret}
\eea
where $g_{hs}(\rr,\rr')$ is the hard sphere pair correlation function, $\sigma$ is the hard sphere diameter.
The primes on velocities denote
scattered values after the collision and 
$\vv_{12}\equiv (\vv_1-\vv_2)$, $\hat \sigma$ is the unit vector directed from 
particle $1$ to particle $2$, and
 $$\vv_1'=\vv_1-(\hat\ss\cdot\vv_{12})\hat\ss$$ 
$$\vv_2'=\vv_2+(\hat\ss\cdot\vv_{12})\hat\ss.$$
The difference from the Boltzmann equation is that the two distribution 
functions are evaluated at different positions.
Such a difference in positions at collision allows the
instantaneous transfer of momentum and energy between two particles.
This collisional transfer is much faster and effective
than the translational transfer and becomes dominant as
density increases.

The salient feature which differentiates RPA and RET is the presence of velocity correlations in the latter
approximation. In fact while it is assumed that the velocities of 
two particles are 
uncorrelated immediately prior to  collision,
but are correlated after they collide, because collision itself generates
correlation.

From eq.(\ref{ret}) we can obtain 
$C^{(1)}_i$ and $C^{(2)}$ once $f(\rr,\vv,t)$ is known.
Interestingly one derives a simple relation for  $C^{(2)}$
in terms of the hydrodynamic variables:
\be
C^{(2)}(\rr,t)=-\nabla_i q_i^{(C)}(\rr,t)-
P_{ij}^{(C)}(\rr,t)\nabla_i u_j(\rr,t)
\ee
where $q_i^{(C)}$ and $P_{ij}^{(C)}$ are the collisional components of
the heat flux vector and of the pressure tensor, respectively.
The two terms in the previous formula account for the transfer of 
thermal energy via collisional transfer and the viscous dissipation
due to the presence of velocity gradients, but such a connection is valid only for the hard sphere case, where
there is no potential energy contribution.

 However, 
it is also possible, as shown in ref. \cite{Melchionna2009},
to obtain explicitly these quantities by approximating in eq. (\ref{ret})
$f(\rr,\vv,t)$ by $f_{loc}(\rr,\vv,t)$, that is, by neglecting $\delta f$ in the integrals.
This approximation
allows to express $C^{(1)}_i(\rr,t)$ and $C^{(2)}(\rr,t)$ by means of 
appropriate convolution integrals involving
finite spatial differences of $n(\rr,t),\uu(\rr,t),$ and $T(\rr,t)$.
It is such a dependence on the hydrodynamic fields which gives a non trivial result
for the transport coefficients. In fact, one finds that in the limit of slowly varying
inhomogeneities the present theory reduces to
to the Longuet-Higgins Pople 
theory of transport coefficients \cite{Longuet}. This level of approximation yields the following
collisional contributions to the shear viscosity
\be
\eta^{(C)}=\frac{4}{15}\sqrt{ \pi}  m v_T \sigma^4 n^2 g_{hs}(\sigma) 
\label{viscosity}
\ee
and to the heat conductivity
\be
\lambda^{(C)}=\frac{2 }{3}\frac{k_B}{m} \sqrt{ \pi} m v_T \sigma^{4}  n^2 g_{hs}(\sigma) .
\label{conductivity}
\ee
To conclude,
in the HS case we obtain a formula for the shear viscosity by
taking into account only the distortion of the singlet distribution function
from the uniform Maxwellian. For more general interactions the potential contribution
to $\eta$ dominates over the one body contribution in liquids at high densities
and one has to consider the distortion of the doublet distribution function
which is completely neglected in the present treatment \cite{morioka}.

\section{Numerical method in a nutshell}
\label{LBE}
%%%%%%%%%%%%%%
This section is included to illustrate how the microscopic description presented
above can be applied in practical situations where the system under scrutiny is
spatially inhomogeneous. In such a case,
the transport equation (\ref{unobrey}) still represents a formidable numerical problem
for the evolution of $f(\rr,\vv,t)$ on a six-dimensional space.
A possible method of solution  is represented by focusing on the
set of partial differential equations for the hydrodynamic moments (\ref{continuity}-\ref{energy}), 
but these require the specification 
of complex boundary conditions and constitutive relations.

An appealing alternative  is   to solve directly eq.(\ref{unobrey}) by using a discretization
of velocities 
on a mesh \cite{shanchen,martys,heluo,shanhe}, a
 technique
 known as Lattice Boltzmann 
Equation (LBE), which deals directly with  the phase space distribution function
and provides a well-tested and robust method of solution \cite{LBgeneral}. 

% In this iterative method of solution, we need only to compute
% $n,\uu,T$ in order to update the local equilibrium distribution
% $f_{loc}$ and the terms $C^{(1)}_i$ and $C^{(2)}$.\cite
%A popular mesh model in three dimensions is provided by the so-called D3Q19 scheme{LBgeneral}, 
In order to illustrate the scheme, we first project the phase space distribution function 
over an orthonormal  basis spanned by the tensorial Hermite polynomials 
$\{H^{(l)}_{\underline\alpha}\}$:
\begin{equation}
f(\rr,\vv,t)=\omega(\vv)\sum_{l=0}^{\infty}\frac{1}{ l! v_{T}^{ 2l} }\phi^{(l)}_{\underline\alpha}(\rr,t)H^{(l)}_{\underline\alpha}(\vv)
\label{sumhermite}
\end{equation}
 where $\omega(\vv)=(2\pi v_{T}^{2})^{-3/2}e^{-\vv^{2}/2v_{T}^{2}}$
 and the Hermite coefficients $\phi^{(l)}_{\underline\alpha}(\rr,t)$, using the orthonormal relation 
 \be
  \int  \omega(\vv)  H^{(l)}_{\underline\alpha}(\vv)  H^{(m)}_{\underline\beta}(\vv)  d\vv=
  (v_T)^{l+m}\delta_{lm}\delta_{\underline\alpha \underline\beta}  
  \ee
  can be obtained by:
 \be
 \phi^{(l)}_{\underline\alpha}(\rr,t)= \int  {f}(\rr,\vv,t) H^{(l)}_{\underline\alpha}(\vv)  d\vv
 \ee
 The exact infinite series  representation  of $\bar{f}(\rr,\vv,t)$ is approximated
 by a function  $\bar{f}(\rr,\vv,t)$  obtained  by retaining in eq.(\ref{sumhermite}) only 
terms  up to  $l = K$, so that
the complete and truncated distributions have the same coefficients up to Hermite order $K$.
At this stage, a   Gauss-Hermite quadrature formula is employed 
in order to evaluate the expansion coefficients, $\phi^{(l)}_{\underline\alpha}(\rr,t)$, 
recognizing that $\bar{f}(\rr,\vv,t)H^{(l)}_{\underline\alpha}(\vv)/\omega(\vv)=p(\rr,\vv,t)$
is a polynomial in $\vv$ with a  degree $\leq 2K$, 
and using the nodes, ${\bf c}_{p}$,  and  the weights  $w_{p}$ of a quadrature of order 
$2G\geq K$, with $p=1,\cdots,Q$.
Using the result
\be
\int d\vv\omega(\vv)p(\rr,\vv,t)\nonumber 
 = \sum_{p=0}^{G}w_{p}p(\rr,{\bf c}_{p},t)
 \ee
one obtains
\begin{eqnarray}
\phi^{(l)}_{\underline\alpha}(\rr,t) =\int d\vv\bar{f}(\rr,\vv,t)H^{(l)}_{\underline\alpha}(\vv)
=\sum_{p=0}^{G}f_{p}(\rr,t)H^{(l)}_{\underline\alpha}({\bf c}_{p})\end{eqnarray}
where 
\be
f_{p}(\rr,t)\equiv \bar{f}(\rr,{\bf c}_{p},t)\frac{w_{p}}{\omega({\bf c}_{p})}
\ee
Notice that $f_{p}(\rr,t)$ contains the same information as the continuous
velocity distribution $\bar{f}(\rr,{\bf v},t)$, so that
the continuous phase space distribution is replaced by a $Q$-dimensional array, 
$\bar{f}(\rr,\vv,t)\rightarrow f_p(\rr,t)$.
The $Q$ nodes connect neighboring
mesh points $\rr$ on a lattice through the discrete velocities ${\bf c}_p$, mirroring the hop of particles
between mesh points, generally augmented by a null vector ${\bf c}_{0}$
accounting for particles at rest.
 The specific form of the lattice
velocities and weights depends on the order of accuracy of the method
and reflects the required Hermite order, as described in the following
and thoroughly discussed in ref. \cite{shanyuanchen}
The LB algorithm exploits the Cartesian mesh to rearrange populations
over spatial shifts corresponding to a first-order fully-explicit
temporal update,
\be
\partial_{t}f_{p}(\rr,t)+\cc_{p}\cdot\partial_{\rr}f_{p}(\rr,t) \simeq
\frac{f_{p}(\rr+\cc_{p}\Delta t,t+\Delta t)-f_{p}(\rr,t)}{\Delta t}
\ee
where $\Delta t$ is the time-step.

Similarly, one expands  the collisional and BGK contributions featuring
in  the right hand side of   eq.\ref{unobrey}  and writes the following expressions:

\be
{\cal B}^{int}(\rr,\vv,t)\rightarrow {\cal B}^{int}_{p}(\rr,t)\equiv
\overline {\cal B}^{int} (\rr,{\bf c}_{p},t)\frac{w_{p}}{\omega(c_{p})}
\ee
and 
\be
{\cal K}(\rr,\vv,t)\rightarrow{\cal K}_{p}(\rr,t)
\equiv\bar{{\cal K}}(\rr,{\bf c}_{p},t)\frac{w_{p}}{\omega(c_{p})}
\ee

Using the specific form of the BGK term  the LB algorithm reads
\begin{equation}
\fl
f_{p}(\rr+{\bf c}_{p} \Delta t ,t+\Delta t)=(1-\frac{\Delta t}{\tau})f_{p}(\rr,t)+
\frac{\Delta t}{\tau}f_p^{eq}(\rr,t)+[{\cal K}_p(\rr,t) 
+{\cal B}_p(\rr,t)] \Delta t \nonumber\\
\end{equation}

The resulting population dynamics of $f_p$ is able to virtually reproduce any target macroscopic
evolutions to high accuracy. In absence of explicit collisional terms, 
and owing to the hyperbolic nature of the evolution equation, 
the Lattice Boltzmann approximates the Navier-Stokes equation in the 
nearly-incompressible limit.
In addition, the LB method is completely flexible
in terms of the mesh spacing $\Delta x$ that can be tuned in order to
resolve the details of the microscopic interactions,
as for the spatial convolution in the collisional terms. 
Therefore, the error introduced in the spatial discretization does not represent
a critical issue.

% \bea
% f_p(\rr+\cc_p,t+1) - f_p(\rr,t) = w_p\sum_{l=0}^K \frac{1}{v_T^{2l}l!} 
% C_{\underline \alpha}^{(l)}(\rr,t) h_{\underline \alpha}^{(l)} (\cc_p)
% +\frac{f_p^{loc}(\rr,t)- f_p(\rr,t)}{\tau_0}
% \nonumber
% \eea

%%%%%%%%%%%%%%%%%%%%%%%%%%%%%%

The specific form of the discrete velocities and weights depends on the order of accuracy of the method
\cite{shanyuanchen}. They are also designed to preserve mass, momentum and local isotropy 
by satisfying the rules $\sum_{p}w_{p}c_{pi}=0$,
$\sum_{p}w_{p}c_{pi}c_{pj}=v_{T}^{2}\delta_{ij}$, $\sum_{p}w_{p}c_{pi}c_{pj}c_{pk}=0$
and $\sum_{p}w_{p}c_{pi}c_{pj}c_{pk}c_{pl}=v_{T}^{4}(\delta_{ik}\delta_{jl}+\delta_{il}\delta_{jk})$,
where $w_p$ is a set of normalized weights and $v_{T}=1/\sqrt{3}$ is the mesh sound speed.

The exact  local equilibrium distribution function contains an infinite number of terms in its Hermite representation, 
but  since  we consider a truncation to second order Hermite polynomials we have 
\begin{equation}
f_{p}^{eq}=w_{p}n(\rr,t)\left[1+\frac{c_{pi}u_{i}(\rr,t)}{v_{T}^{2}}+\frac{(c_{pi}c_{pj}-v_{T}^{2}\delta_{ij})u_{i}(\rr,t)u_{j}(\rr,t)}{2v_{T}^{4}}\right]\end{equation}
corresponding to a low-Mach ($O[Ma^{3}]$) expansion of the
local Maxwellian. On the other hand, by keeping only
 the momentum component of the collisional contribution, ${\bf C}^{(1)}(\rr,t)$ , one writes 
 the following   discretized form:
\begin{equation}
{\cal K}_{p}=-w_{p}\frac{1}{m}\left[\frac{c_{pi}C_{i}^{(1)}(\rr,t)}{v_{T}^{2}}+\frac{(c_{pi}c_{pj}-v_{T}^{2}\delta_{ij})u_{i}(\rr,t)C_{j}^{(1)}(\rr,t)}{v_{T}^{4}}\right]
\end{equation}
Similarly, the contribution arising from heat conduction, $C^{(2)}(\rr,t)$, could be included within 
the same framework. However, a thermal version of the LB scheme requires inclusion up to
third or fourth order Hermite polynomials.

Once the populations $f_p$ are known, they are used to compute hydrodynamic moments, 
entering both the equilibrium and in sampling the macroscopic evolution.
The fluid density, momentum current and local temperature read
\bea
n(\rr,t) &=& \sum_p f_p(\rr,t) \nonumber \\
n(\rr,t) \uu(\rr,t) &=& \sum_p f_p(\rr,t) \cc_p \nonumber \\
\frac{3 k_B}{m} n(\rr,t) T(\rr,t) &=& \sum_p f_p(\rr,t) \cc_p^2
\eea
A Chapman-Enskog analysis shows that the relaxation time $\tau$ is related to the kinetic component 
of the shear viscosity via 
\begin{equation}
\eta^{(K)}=n m v_T^2(\tau-\frac{\Delta t}{2})
\label{visctau}
\end{equation}
i.e. the physical value, found in eq.(\ref{viscosityk}), subtracted by a contribution of numerical origin.

%In some applications it is necessary to compute ${\bf C}_i^{(1)}(\rr,t)$  and ${\bf C}^{(2)}(\rr,t)$ 
%from spatial convolutions of the inhomogeneous pair correlation function 
%with the hydrodynamic fields.  To this purpose we evaluated the pair correlation
%function $g(\rr,\rr)$
For hard spheres,
the radial distribution function $g(\rr,\rr')$ appearing in the evaluation of
${\bf C}_i^{(1)}(\rr,t)$  and ${\bf C}^{(2)}(\rr,t)$  associated with the RET approximation is
obtained according to the 
Fischer and Methfessel construction \cite{fishmet}.  At first, one defines a
coarse-grained density $\bar n (\rr,t)$ via a uniform smearing over a
sphere of radius $\sigma/2$, and the coarse-grained packing fraction
is $\bar \eta (\rr,t) = \pi\sigma^3 \bar n(\rr,t)/6$. The
actual pair correlation, $g_{hs}(\rr,\rr+\ss)$, is
replaced by its equilibrium value at the given smoothed density   
$$
g_{hs}(\rr,\rr+\ss) \simeq
[1-\bar\eta(\rr+\ss/2)/2+\bar\eta^2(\rr+\ss/2)/4]/[1-\bar\eta(\rr+\ss/2)]^3.
$$
In order to evaluate the surface integrals, we choose $\sigma$ to be an
even multiple of the lattice spacing and employ a $18$-point
quadrature over a spherical surface \cite{abramovitzstegun}.  With
this choice, the elements arising from $g_{hs}$ and the hydrodynamic
moments are taken from $6$ on-lattice quadrature points while the
elements arising from the remaining $12$ off-lattice points are
constructed via a linear interpolation from the surrounding on-lattice
elements.

The numerical method has been tested to determine the shear viscosity of a uniform system
and  an excellent agreement was found with theoretical prediction eq.(\ref{viscosity}).
In addition we have considered the flow of an hard sphere system  induced in a slit by the presence
of a uniform field parallel to the walls. The observed velocity profiles had the expected parabolic 
behavior for wall distances sufficiently larger than the hard sphere diameter and moderately low
densities, but displayed pronounced oscillations for larger densities and the average streaming 
velocity decreased as a consequence of mutual steric hindrance among particles.

%%%%%%%%%%%%%%%%%%%%%%%%%%%%%%%%%%%%%%%%%%%%%%%%%

Regarding the stability of the method, 
LB is subjected to numerical instability whenever the flow velocity becomes larger 
than a certain threshold, function of the relaxation time, the wave number  \cite{shanyuanchen}
and the stiffness of intermolecular forces. However, the stability range can always be widened 
by taking a smaller $\Delta x$. In fact, a generic upper bound for the variation of
populations due to the forcing term is $\delta f/f\sim f_{p}/w_{p}\sim\Delta t{\cal K}_{p}\ll1$.

The advantages of the LB method over the solution of the coupled hydrodynamic equations
are evident:
a) the convective term $\uu(\rr,t)\cdot {\bf \nabla}$ does not need to be computed explicitly;
b) the solution of the Poisson equation for the pressure is not needed;
c) the boundary conditions are handled in an easier way, for instance by imposing 
simple collision rules on the local velocity the distribution at the boundary, such as bounce-back
in order to reproduce a no-slip condition;
d) a relatively small number of  velocity and spatial mesh points are sufficient for many
purposes;
e) the  two-body interparticle potential can be quite arbitrary and in the  case of a slowly
varying attractive tail it can be  modeled via the RPA.
Such an approximation, although not contributing to the shear viscosity,
it allows to study phase coexistence.
%%%%%%%%%%%%%%%%%%%%%%%%%%%
\section{ Conclusions}
\label{conclusions}

Starting from a microscopic level we have obtained a governing equation
for $f(\rr,\vv,t)$ describing both equilibrium structural properties
and  transport properties.
The essential difference between  DDFT and the present method  
consists in the fact in the first the
dynamical evolution of the fluid system is treated 
only in configurational space and the fundamental variable  is the one 
particle density, whereas in the latter the dynamics is studied 
in phase space and the corresponding quantity is the
phase space density distribution.
 The use of such a distribution enables
us to consider the microscopic details of the molecular collisions
and analyze transport properties on a quantitative basis. 

The DDFT equation can be obtained  as a particular case, when the heat-bath term
is externally imposed and has a viscous character forcing the velocity distribution 
to attain its global equilibrium form in a time much shorter than the time needed to
reach the configurational equilibrium.
As a result, DDFT does not describes sound waves in liquids, but only diffusive modes.
 In the case of molecular fluids
instead, the time scales
characterizing the equilibration times of the hydrodynamic fields can be comparable and 
diverge as the wavelength of the modes tends to infinity,
as a result of the translational invariance and absence of a fixed 
solvent. The  kinetic
equation (\ref{unobrey}) by construction reproduces this aspect and
the mechanism restoring the global equilibrium in the fluid is purely
due to the presence of gradients in the hydrodynamic fields and not
to the fields themselves. 
Hydrodynamic versus non-hydrodynamic modes splitting proves a convenient
route. 
As far as similarities are concerned,
both methods use the same equilibrium structural information as their key ingredient
and the same adiabatic treatment of two particle correlation necessary to decouple the
one particle from the multiparticle dynamics. 

Other popular approaches within the LB framework have been put forward to treat inhomogeneous
fluids, but they differ from ours in the derivation  of the interaction terms. Most of these methods,
such as the Swift-Osborne-Yeomans \cite{Yeomans} and the effective pseudo potential  methods  
by Shan-Chen \cite{shanchen},
do not utilize a microscopic, but rather start from a coarse grained level and assume
the validity of a local thermodynamics.   
In our opinion, we find  rewarding to be able to derive in  unified way both DDFT
and transport equation and to discriminate them on the basis of the thermostatting mechanism.

The major limit of the present treatment is the assumption underlying
the Boltzmann equation that particles are uncorrelated before
collisions, resulting in a markovian dynamics.  In the case of hard
spheres, the RET approximation represents an improvement over the
Boltzmann treatment by taking into account the non local character of
the momentum and energy transfer, but without incorporating the time
correlations which are responsible for memory effects.  Even the RET
approach is local in time with no memory of the past history.  The
deficiency of the theory becomes more acute when treating soft
potentials, where the interaction is not instantaneous as in the
hard-sphere case.  For continuous interaction, it is not possible to
consider the dynamics as composed of two separate stages, namely
instantaneous collisions followed by free streaming trajectories, as
in the case of hard spheres. The interaction have a finite duration
and a second interaction can take place before the first one is
completed.
Some authors have modeled the different regions of the
 attractive potential tails by means of sequence of square-well constant 
potentials 
 so that the RET formalism could be applied \cite{Stell}. 
%Such a program has been carried out in the case
%of homogeneous attractive fluids and their transport coefficient have been shown to be in agreement
%with numerical simulations. 
However, this approach inherits from the hard-sphere model the
impulsive character of the interactions and therefore fails to reproduce 
memory effects. 
%%%%%%%%%%%%%
Alternatively, it has been observed \cite{RiceAllnatt}  that  the motion of the
particles  can be regarded as  the combination of continuous momentum changes, assimilated 
to a  Brownian motion in an average attractive field, and  quasi-instantaneous collisions with 
large  momentum binary exchanges, described as hard core collisions. 
Following these lines it seems possible to extend the present approach to the treatment of soft potentials.

%%%%%%%%%%%

%\acknowledgments

%\newpage
%%%%%%%%%%%%%%%%%%%%%%%%%%%%%%%%%%%%%%%%%%5

\end{document}